\documentclass[12pt]{article}
\pdfoutput=1
\usepackage{amsmath}
\usepackage[dvips]{graphicx}
\DeclareGraphicsExtensions{.png,.pdf}

\usepackage{color}
\usepackage{pstricks}
\input amssym

\def\red#1{{\color{red} #1}}

\begin{document}

\def\prg#1{\par\medskip\noindent{\bf #1}}  \def\ra{\rightarrow}
\newcounter{nbr}
\def\note#1{\bitem\vspace{-5pt}\addtocounter{nbr}{1}
            \item{} #1\vspace{-5pt}
            \eitem}
\def\lra{\leftrightarrow}              \def\Ra{\Rightarrow}
\def\nin{\noindent}                    \def\pd{\partial}
\def\dis{\displaystyle}                \def\Lra{{\Leftrightarrow}}
\def\grl{{GR$_\Lambda$}}               \def\vsm{\vspace{-8pt}}
\def\cs{{\scriptstyle\rm CS}}          \def\ads3{{\rm AdS$_3$}}
\def\Leff{\hbox{$\mit\L_{\hspace{.6pt}\rm eff}\,$}}
\def\bull{\raise.25ex\hbox{\vrule height.8ex width.8ex}}
\def\ric{{Ric}}                        \def\tmgl{\hbox{TMG$_\Lambda$}}
\def\Lie{{\cal L}\hspace{-.7em}\raise.25ex\hbox{--}\hspace{.2em}}
\def\sS{\hspace{2pt}S\hspace{-0.83em}\diagup}
\def\hd{{^\star}}                      \def\dis{\displaystyle}
\def\mb#1{\hbox{{\boldmath $#1$}}}     \def\kn#1{\hbox{KN$#1$}}
\def\ul#1{\underline{#1}}              \def\phb{\phantom{\Big|}}
\def\nb{~\marginpar{\bf\Large ?}}      \def\ph{\phantom{xxx}}

\def\hook{\hbox{\vrule height0pt width4pt depth0.3pt
\vrule height7pt width0.3pt depth0.3pt
\vrule height0pt width2pt depth0pt}\hspace{0.8pt}}
\def\inn{\hook}
\def\first{\rm (1ST)}  \def\second{\hspace{-1cm}\rm (2ND)}
\def\ppl{{pp${}_\Lambda$}}

\def\G{\Gamma}        \def\S{\Sigma}        \def\L{{\mit\Lambda}}
\def\D{\Delta}        \def\Th{\Theta}       \def\Ups{\Upsilon}
\def\a{\alpha}        \def\b{\beta}         \def\g{\gamma}
\def\d{\delta}        \def\m{\mu}           \def\n{\nu}
\def\th{\theta}       \def\k{\kappa}        \def\l{\lambda}
\def\vphi{\varphi}    \def\ve{\varepsilon}  \def\p{\pi}
\def\r{\rho}          \def\Om{\Omega}       \def\om{\omega}
\def\s{\sigma}        \def\t{\tau}          \def\eps{\epsilon}
\def\nab{\nabla}      \def\btz{{\rm BTZ}}   \def\heps{{\hat\eps}}

\def\bR{\bar{R}}      \def\bT{\bar{T}}     \def\hT{\hat{T}}
\def\tG{{\tilde G}}   \def\cF{{\cal F}}    \def\cA{{\cal A}}
\def\cL{{\cal L}}     \def\cM{{\cal M }}   \def\cE{{\cal E}}
\def\cH{{\cal H}}     \def\hcH{\hat{\cH}}  \def\cT{{\cal T}}
\def\hA{\hat{A}}      \def\hB{\hat{B}}     \def\hK{\hat{K}}
\def\cK{{\cal K}}     \def\hcK{\hat{\cK}}  \def\cT{{\cal T}}
\def\cO{{\cal O}}     \def\hcO{\hat{\cal O}} \def\cV{{\cal V}}
\def\tom{{\tilde\omega}}  \def\cE{{\cal E}} \def\bH{\bar{H}}
\def\cR{{\cal R}}    \def\hR{{\hat R}{}}   \def\hL{{\hat\L}}
\def\tb{{\tilde b}}  \def\tA{{\tilde A}}   \def\hom{{\hat\om}}
\def\tT{{\tilde T}}  \def\tR{{\tilde R}}   \def\tcL{{\tilde\cL}}
\def\he{{\hat e}}    \def\hom{{\hat\om}}   \def\hth{\hat\theta}
\def\hxi{\hat\xi}    \def\hg{\hat g}       \def\hb{{\hat b}}
\def\tH{{\tilde H}}  \def\tV{{\tilde V}}   \def\bA{\bar{A}}
\def\bV{\bar{V}}     \def\bxi{\bar{\xi}}
\def\knl{\text{KN}$(\l)$}   \def\mknl{\text{KN}\mb{(\l)}}
\def\bPhi{\bar\Phi}
\def\chm{\checkmark}                \def\chmr{\red{}}
\vfuzz=2pt 
\def\nn{\nonumber}
\def\be{\begin{equation}}             \def\ee{\end{equation}}
\def\ba#1{\begin{array}{#1}}          \def\ea{\end{array}}
\def\bea{\begin{eqnarray} }           \def\eea{\end{eqnarray} }
\def\beann{\begin{eqnarray*} }        \def\eeann{\end{eqnarray*} }
\def\beal{\begin{eqalign}}            \def\eeal{\end{eqalign}}
\def\lab#1{\label{eq:#1}}             \def\eq#1{(\ref{eq:#1})}
\def\bsubeq{\begin{subequations}}     \def\esubeq{\end{subequations}}
\def\bitem{\begin{itemize}}           \def\eitem{\end{itemize}}
\renewcommand{\theequation}{\thesection.\arabic{equation}}

\title{Velocity memory effect without soft particles}

\author{B. Cvetkovi\'c and D. Simi\'c\footnote{
        Email addresses: {\tt cbranislav@ipb.ac.rs,
                          dsimic@ipb.ac.rs}}\\
Institute of Physics, University of Belgrade \\
                      Pregrevica 118, 11080 Belgrade, Serbia}
\date{\today}
\maketitle

\begin{abstract}
We study the behavior of geodesics in the plane-fronted wave background of the three-dimensional  gravity with propagating torsion, which possesses only {\it massive} degrees of freedom.
 We discover the {\it velocity memory effect},  in contrast to the current belief that its existence  is due to the  presence of soft particles.
\end{abstract}

\section{Introduction}
\setcounter{equation}{0}
Memory effect for gravitational waves was first discovered by Zeldovich and Polnarev \cite{1} and got its name from Braginsky and Grishchuk \cite{2}. The conclusion of \cite{1,2} is that massive test particles, initially at rest, will suffer permanent displacement after the passage of gravitational wave. For this reason  this displacement is called memory effect.

The memory effect \cite{1,2} is described in linear approximation. A nonlinear contribution to the memory effect is discovered in reference \cite{3}, for less technical derivation see \cite{4}.

In the recent years, we have witnessed a great new discoveries connecting asymptotic symmetries, soft theorems, and displacement memory effect \cite{5}. This line of reasoning applied to the black holes \cite{5'} offers new insights into black hole physics.

Permanent displacement implies that relative velocity of  massive test particles is zero. This conclusion is questioned in references \cite{6,7}, where velocity memory effect is derived on the contrary to displacement memory effect. The main result of \cite{6,7} is that passage of gravitational wave will be encoded not in the permanent displacement  but in the nonzero relative velocity of test masses.
For recent development on velocity memory effect see, \cite{8}, where, among other things, authors concluded that velocity memory effect is connected with soft gravitons.

The goal of this paper is to investigate if   there is a memory effect for gravitational waves within the framework of Poincar\'e gauge theory \cite{x3,x4,x5}. We shall consider the solutions of three-dimensional (3D)  gravity with propagating torsion \cite{9,10},  a theory
in which all modes are massive. If the soft modes
are the ones that cause memory effect (according to the already mentioned conclusions in \cite{8}) there should be no memory effect in this theory.

Let us note that 3D general relativity (GR) is a topological theory and there are consequently no  gravitational wave solutions in vacuum. The gravitational waves with torsion in 3D  are solutions in which the metric function {\it crucially} depends on torsion \cite{10}, in the sense that in the absence of torsion, the metric function becomes trivial and the wave solution "disappears". This offers us an interesting opportunity to study the effects of torsion already at the level of geodesic motion of spinless particles.

The paper is organized  as follows. First, we review the theory of gravity under consideration and its gravitational pp wave solutions. Next, we derive the geodesic equations in this pp wave space-time. Thereafter,  we investigate the solutions of this equations. Unfortunately, the geodesic equations are not analytically solvable except in a very special case, which we have also analyzed; thus, we solved the geodesic equations numerically for some characteristic choices of the coefficients which appear in the gravitational wave solutions.

Our conventions are as follows. The Latin indices $(i, j, ...)$ refer to
the local Lorentz (co)frame and run over $(0,1,2)$, $b^i$ is the tetrad
(one form), $h_i$ is the dual basis (frame), such that $h_i\inn b^k=\d^i_k$;
the volume three form is $\heps=b^0\wedge b^1\wedge b^2$, the Hodge
dual of a form $\a$ is $\hd\a$, with $\hd 1=\heps$, totally antisymmetric
tensor is defined by $\hd(b_i\wedge b_j\wedge b_k)=\ve_{ijk}$
and normalized to $\ve_{012}=+1$; the exterior product of forms is
implicit.

\section{ Riemannian pp waves}
\setcounter{equation}{0}

In this section, we give an overview of Riemannian 3D pp waves. For details see \cite{9}.

\subsection{Geometry}

The metric of  pp waves can be written as
\bsubeq\lab{2.1}
\be
ds^2=du(Sdu+dv)-dy^2\, ,     \lab{2.1a}
\ee
where
\be
S=\frac12H(u,y)\, .
\ee
\esubeq
Next, we choose the tetrad field (coframe) in the form
\bsubeq\lab{2.2}
\be
b^0:=du\, ,\qquad b^1:=Sdu+dv\, ,\qquad
b^2:=dy\,  ,
\ee
so that $ds^2=\eta_{ij}b^i\otimes b^j$, where $\eta_{ij}$ is the half-null
Minkowski metric
$$
\eta_{ij}=\left( \ba{ccc}
             0 & 1 & 0  \\
             1 & 0 & 0   \\
             0 & 0 & -1
                \ea
       \right)\, .
$$
The corresponding dual frame $h_i$ is given by
\be
h_0=\pd_u-S\pd_v\, ,\qquad h_1=\pd_v\, ,
\qquad h_2=\pd_y\,  .
\ee
\esubeq
For the coordinate $x^\a=y$ on the wave surface, we have
$$
x^c=b^c{_\a}x^\a=y\, ,\qquad
\pd_c=h_c{^\a}\pd_\a=\pd_y\, ,
$$
where $c=2$.

Starting from the general formula for the Riemannian connection one form,
$$
\om^{ij}:=-\frac{1}{2}\Bigl[h^i\inn db^j-h^j\inn db^i
                           -(h^i\inn h^j\inn db^k)b_k\Bigr]\, ,
$$
one can find its explicit form; for $i<j$, its nonvanishing component  reads as
\bsubeq\lab{2.3}
\bea
\om^{12}=-\pd_y S\,b^0\,.
\eea
Introducing the notation $i=(A,a)$, where $A=0,1$ and $a=2$, one can
rewrite $\om^{ij}$ in a more compact form as follows:
\bea
&&\om^{Ac}=k^Ab^0\pd^c S\, ,
\eea
\esubeq
where $k^i=(0,1,0)$ is a null propagation vector, $k^2=0$.

The above connection defines the Riemannian curvature
\bsubeq
\be
R^{ij}=2b^0k^{[i}Q^{j]}\, ,
\ee
where
\bea
&&Q^2=\pd_{yy}Sb^2
\eea
\esubeq
The Ricci one form $\ric^i:=h_m\inn\ric^{mi}$ is given by
\bsubeq
\bea
&&\ric^i=b^0k^i Q\, ,\qquad Q=h_c\inn Q^c=\frac{1}{2}\pd_{yy}H,         \lab{2.5}
\eea
and the scalar curvature vanishes
\be
R=0\,.
\ee
\esubeq
\subsection{Dynamics}

\subsubsection{pp waves in GR}

Starting with the action $I_0=-\int d^4 x(a_0R+2\L_0)$, one can derive the
GR field equations in vacuum,

\be
2a_0G^n{_i}=0\, ,
\ee
where $G^n{_i}$ is the Einstein tensor.
As a consequence, the metric function $H$ must obey
\be
\pd_{yy}H=0\, .                      \lab{2.8}
\ee
However, the solution of this equation is trivial,
$$
H=C(u)+yD(u)\,,
$$
since the corresponding radiation piece of curvature vanishes.
\section{pp waves with torsion}\label{sec3}
\setcounter{equation}{0}

\subsection{Geometry of the ansatz}

We assume that the
form of the triad field \eq{2.2} remains unchanged, whereas the connection
is
\bsubeq\lab{3.3}
\bea
&&\om^{ij}=\tom^{ij}+\frac{1}{2}\ve^{ij}{_m}k^m k_n b^n G\, ,   \lab{3.3a}\\
&&G:=S'+K\, .                                    \lab{3.3b}
\eea
\esubeq
Here, the new term $K=K(u,y)$ describes the effect of torsion, as follows:
\be
T^i:=\nab b^i=\frac{1}{2}Kk^ik_m\hd b^m\, .                  \lab{3.4}
\ee
The only nonvanishing irreducible piece of $T^i$ is its tensorial piece
$$
{}^{(1)}T^i=T^i\, ,
$$
while the curvature is
\bea
&&R^{ij}=\ve^{ijm}k_m k^n\hd b_n\, G' \, ,          \nn\\
&&\ric^i=\frac{1}{2}k^ik_mb^m G'\,,                   \nn\\
&&R=0\, .                                                    \lab{3.5}
\eea
The nonvanishing irreducible components of the curvature $R^{ij}$ are
$$
{}^{(4)}R^{ij}=\frac{1}{2}\ve^{ijm}k_m k^n\hd b_n\, pG'\,.
$$
and the quadratic curvature invariant vanishes
$R^{ij}\,{}^*R_{ij}=0$. For details on irreducible decomposition of
torsion and curvature see \cite{x23}.

The geometric configuration defined by the triad field \eq{2.2} and the
connection \eq{3.3} represents a generalized gravitational plane-fronted
wave of \grl, or the \emph{torsion wave} for short.

\subsection{Massive torsion waves}
\setcounter{equation}{0}

The field equations take the following form \cite{10}:
\bea
&&a_0G'-a_1K'=0\, ,\qquad \L=0\, ,                              \nn\\
&&K''+m^2K=0\, ,\qquad m^2=\frac{a_0(a_1-a_0)}{b_4 a_1}\,,      \lab{4.1}
\eea
with $G=S'+K$ and $S=H/2$. The solution has a simple form,
\bea
&&K=A(u)\cos my+B(u)\sin my\, ,                                 \nn\\
&&\frac{1}{2}H=
    \frac{a_1-a_0}{a_0 m}(A(u)\sin my-B(u)\cos my)+h_1(u)+h_2(u)y\, . \lab{4.2}
\eea
Disregarding the integration ``constants" $h_1$ and $h_2$, the metric and
the torsion functions, $H$ and $K$, become both periodic in $y$.
In the absence of torsion  the metric function becomes trivial. This is an expected result
since 3D general relativity is a theory which possesses no propagating degrees of freedom.

The vector field $k=\pd_v$ is the Killing vector for both the metric and
the torsion; moreover, it is a null and covariantly constant vector field.
This allows us to consider the solution \eq{4.2} as a generalized pp wave.


\section{Geodesic motion}
\setcounter{equation}{0}

In this section, we shall examine the {\it geodesic motion} of particles in the field of the massive wave with torsion.

At first sight that might be puzzling, since the motion of spinless particles is not affected by torsion \cite{x5,y1}, and in the
gravitational field, they follow geodesic lines, which are influenced by the Riemannian connection depending on the metric. However, gravitational waves with torsion in 3D are interesting solutions, which are intrinsically  different from  the well-known spherically symmetric (static or stationary) solutions of  Poincar\'e gauge theory \cite{y2} (for review, see \cite{y3}). The  metric of these solutions is "independent" of torsion in the sense that it represents Schwarzschild (or Schwarzschild anti de Sitter, Kerr etc.)  metric and the motion of spinless particles is not affected by the presence of torsion. However, for the gravitational wave solution \eq{4.2}, metric crucially depends on torsion as we noted in the previous section. This offers us an interesting
opportunity to study the effects of torsion already at the  level of geodesic motion.
\prg{Christoffel  connection.} The nonvanishing components of Christoffel (torsion free) connection are given by
\bea
&&\tilde{\G}^v{}_{uu}=\frac12\pd_u H\,,\nn\\
&&\tilde\G^v{}_{uy}=\frac12H'\,,\qquad \tilde\G^v{}_{yu}=\frac12H'\,,\nn\\
&&\tilde\G^y{}_{uu}=\frac12H'\,.
\eea
Let us mention that nontrivial contribution to metric function and consequently Christoffel connection stems from the presence of torsion.
\prg{Geodesic equations.} The geodesic equation for $u$ takes the expected form
\bea
\frac{d^2u}{d\l^2}=0\,.
\eea
Therefore, without the loss of generality, we can assume $u\equiv \l$.

The equation for $y$ is given by
\bsubeq
\be
\ddot{y}+\frac 12 H'=0\,,
\ee
or more explicitly
\be
\ddot{y}+\frac{a_1-a_0}{a_0}(A\cos my+B\sin my)=0\,.\lab{4.3b}
\ee
\esubeq
Finally, the equation for $v$ reads as

\be
\ddot{v}+\frac12\pd_u{H}+H'\dot y=0\,,\lab{4.4}
\ee

In the special case, when $H$ does not explicitly depend on $u$, the  equation \eq{4.4} can be integrated as
$$
\dot v+Hy=C\,,
$$
where $C$ is a integration constant.
Hence, consequently, we get:
\be
v=\int (C-Hy)du\,.\lab{4.5}
\ee
\subsection{Exact solutions}

Interestingly,  the geodesic equations admit the existence of exact solutions in particular cases.
The simplest case is when $A(u)$ and $B(u)$ are constants.
In that case \eq{4.3b} can be rewritten in the form
\bea
\frac 12\frac{d{\dot y^2}}{dy}+\frac{a_1-a_0}{a_0}\left(A\cos my+B\sin my\right)=0\,.\nn
\eea
If we impose {\it initial conditions}
\be
y(0)=0\,,\qquad \dot y(0)=0\,,
\ee
by integrating the previous equation, we obtain
$$
\frac12\dot y^2+\frac{a_1-a_0}{a_0m}\left(A\sin my-B(\cos my-1)\right)=0\,,
$$
or equivalently:
\bea
&&\frac{dy}{\sqrt{\bar A\sin my-\bar B(\cos my-1)}}=du\,,\nn\\
&&\bar A:=\frac{2(a_0-a_1)}{a_0m}A\,,\qquad
\bar B:=\frac{2(a_0-a_1)}{a_0m}B\,,\nn
\eea
which after integration yields the following equation for $y$:
\bea
&&\frac{4i\sqrt{2\bar A\bar B-(\bar A^2+\bar B^2)\frac{\cos\frac{my}2}{\sin^2\frac {my}4}}}
{m\sqrt{\bar B+\sqrt{\bar A^2+\bar B^2}}\sqrt{\bar A\sin my-\bar B(\cos my-1)}}\sin\frac{my}2\sqrt{\tan\frac{my}4}\times\nn\\
&&\times F\left(\left.i{\rm Arcsinh}\left(\frac{\sqrt{\frac{\bar B+\sqrt{\bar A^2+\bar B^2}}{A}}}{\sqrt{\tan\frac{my}4}}\right)\right|
-\frac{\bar A^2+2\bar B(\bar B-\sqrt{\bar A^2+\bar B^2}}{\bar A^2}\right)=u\,,
\eea
where $F(\phi|k)$ represents the elliptic integral of the first kind \cite{x19}.

The choice $\bar A\neq 0$, $B(u)=0$ yields the following exact solution for $y(u)$:
\bea
y(u)=-\frac{2{\rm am}\left(\frac12\sqrt{\bar A}mu\left.|2\right.\right)}{m}\,,
\eea
where ${\rm am}(z|m)$ is a Jacobi amplitude function.
$H$ does not explicitly depend on $u$ we get that $v$ is given by the expression \eq{4.5}.

The characteristic plots for particle position $y$ and velocity $\dot y$  (for $m=2$, $\bar A=1$) are shown in Fig. 1 and Fig 2, respectively.
\begin{figure}
\centering
\includegraphics[height=4cm]{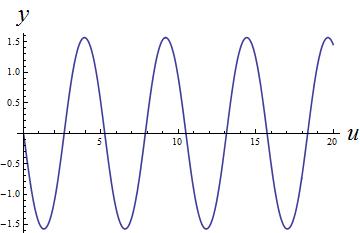}
\caption{The plot for the particle position in units  $m=2$, for $\bar A=1$}
\end{figure}
\begin{figure}
\centering
\includegraphics[height=4cm]{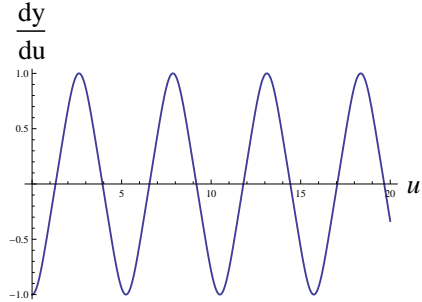}
\caption{The plot for the particle velocity in units  $m=2$, for $\bar A=1$}
\end{figure}
\newpage
\subsection{Velocity memory effect}

The velocity memory effect is present in the case when functions $A(u)$ and $B(u)$ vanish for large $u$.

\prg{Shockwave case.} In the shock wave case, when functions $A(u)=0$ and $B(u)$  vanish exponentially, for example $B(u)\sim e^{-(u-10)^2}$ numerical solutions of the geodesic equations
lead to the plots for the particle position $y$ and $v$  shown in the Figure 3 and velocity $\dot y$ and $\dot v$ shown in the Figure 4.
 \begin{figure}
\centering
\includegraphics[height=4cm]{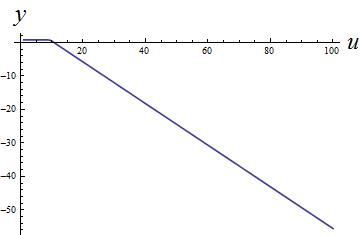}\hspace{1cm}\includegraphics[height=4cm]{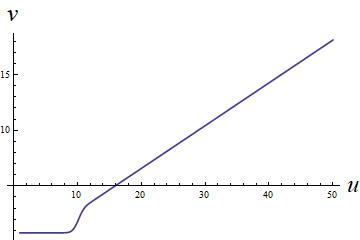}
\caption{The plot for the particle position $y$ and $v$ in units  $m=1$, for $\bar B=-e^{-(u-10)^2}$}
\end{figure}
\begin{figure}
\centering
\includegraphics[height=4cm]{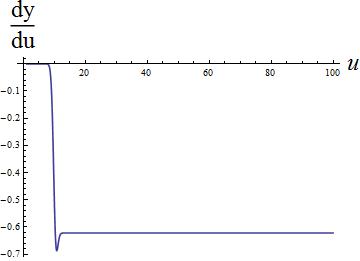}\hspace{1cm}\includegraphics[height=4cm]{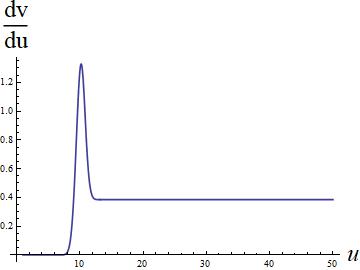}
\caption{The plot for the particle velocity $\dot y$ and $\dot v$ in units  $m=1$, for $\bar B=-e^{-(u-10)^2}$}
\end{figure}
\prg{Slow fall off.} In the case when $A(u)=0$ and $B(u)\sim 1/u$ numerical solutions lead to the following plots for the particle position $y$ and $v$ shown in the figure 5 and velocity $\dot y$ and $\dot v$ shown in the Figure 6:
\begin{figure}
\centering
\includegraphics[height=4cm]{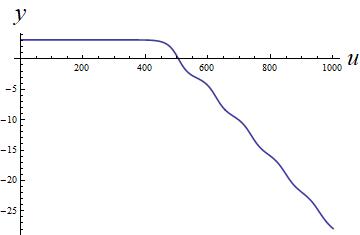}\hspace{1cm}\includegraphics[height=4cm]{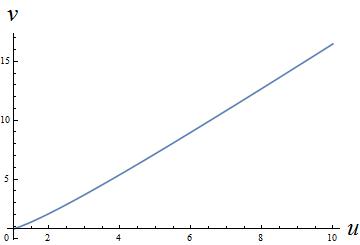}
\caption{The plot for the particle position $y$ and $v$ in units  $m=1$, for $\bar B=-1/u$}
\end{figure}
\begin{figure}
\centering
\includegraphics[height=4cm]{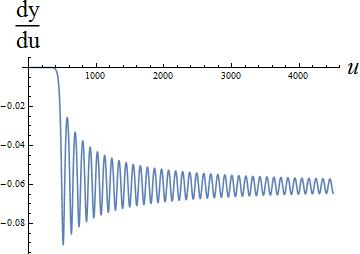}\hspace{1cm}\includegraphics[height=4cm]{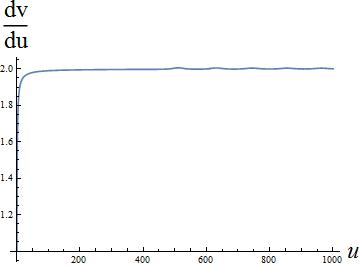}
\caption{The plot for the particle velocity $\dot y$ and $\dot v$ in units  $m=1$, for $\bar B=-1/u$}
\end{figure}
\newpage

\section{Discussion}
We studied the geodesic motion in asymptotically flat pp wave space-time and we discovered the presence of velocity memory effect.
The effect is present for the very fast falloff of the gravitational wave, as well as for the slow one.
Analysis of this paper provides the first example of memory effect for gravitational waves with torsion.
We demonstrated that torsion waves lead to the memory effect  same as the torsion-less waves do.
This is, also the first account of the memory effect in three dimensions to authors knowledge.
It would be interesting to see is
there a connection to BMS$_3$ symmetry.
Because theory has no massless modes, without a doubt, we can conclude that there can be no soft particles responsible for the memory effect. Consequently, the belief that soft particles are responsible for the velocity memory effect is demonstrated to be incorrect. For the related work on massive gravity, see \cite{17}.

Intuitively, we can say that memory effect is due to energy transfer. Passing gravitational wave transfers energy to the test particle which after the passage of the gravitational wave continues to move with constant velocity, in which intensity is dictated by the amount of energy transferred. Looking at the memory effect in this way we conclude that displacement memory effect is not possible, except, maybe, in some special cases where the total amount of transferred energy would be zero.  To make this intuitive discussion precise it is required to define energy in asymptotically flat space-times in a satisfying manner; this is left for further investigation.

The theory we considered is three-dimensional, while the four-dimensional case is realistic and relevant for applications.
The next step in investigation is to study geodesic motion for massive gravitational waves with torsion in four dimensions. The metric of the gravitational waves with torsion
in 4D has a nontrivial contribution stemming from the tensorial component of torsion \cite{y4} as in 3D, which  affects geodesic motion.  Consequently, it is expected that
in 4D we shall obtain the velocity memory effect similar to the one noticed in 3D case.   Also, there is a possible difference compared to the memory effect in general relativity,  which, in principle, may be observable.   This will be the possible experimental setup for detection of torsion.

\section*{Acknowledgments}

This work was partially supported by the Serbian Science Foundation under Grant No. 171031.


\begin{thebibliography}{99}
\bibitem{1} Ya. B. Zeldovich and A. G. Polnarev, Radiation of gravitational waves by a cluster of
superdense stars, Astron. Zh. {\bf 51}, 30 (1974) [Sov. Astron. {\bf 18} 17 (1974)].

\bibitem{2} V. B. Braginsky and L. P. Grishchuk, Kinematic resonance and the memory effect in free mass
gravitational antennas, Zh. Eksp. Teor. Fiz. {\bf 89} 744 (1985) [Sov. Phys. JETP {\bf 62}, 427 (1985)].

\bibitem{3} D. Christodoulou, Nonlinear Nature of Gravitation and Gravitational Wave Experiments,
Phys. Rev. Lett. {\bf 67} 1486 (1991).

\bibitem{4} K. S. Thorne, Gravitational-wave bursts with memory: The Christodoulou effect, Phys.
Rev. D {\bf 45} 520, (1992).


\bibitem{5} Temple He, V. Lysov, P. Mitra and A. Strominger, BMS supertranslations and Weinbergs soft graviton theorem, JHEP {05} (2015) 151;\\
A. Strominger and A. Zhiboedov, Gravitational memory, BMS supertranslations and soft theorems, JHEP {\bf 01} (2016) 086.

\bibitem{5'}S. W. Hawking, M. J. Perry and A. Strominger, Soft Hair on Black Holes, Phys. Rev. Lett {\bf 116}, 231301 (2016).

\bibitem{6} H. Bondi and F. A. E. Pirani, Gravitational waves in general relativity III. Exact plane waves, Proc. R. Soc.  A {\bf 251}, 519 (1959).

\bibitem{7} L. P. Grishchuk and A. G. Polnarev, Gravitational wave pulses with velocity coded memory,
Zh. Eksp. Teor. Fiz. {\bf 96} (1989) 1153 [Sov. Phys. JETP {\bf 69} (1989) 653].

\bibitem{8}P.-M Zhang, C. Duval, G.W. Gibbons and P.A. Horvathy, The memory effect for plane gravitational waves, Phys. Lett. B 743 {\bf 772} (2017);\\
P. M Zhang, C. Duval, G.W. Gibbons and P.A. Horvathy, Soft gravitons and the memory effect for plane gravitational waves, Phys. Rev. D {\bf 96},
 064013 (2017) no.6;\\
 P. M. Zhang, C. Duval, G.W.~Gibbons and P. A. Horvathy,
  Velocity memory effect for polarized gravitational waves, JCAP {\bf 05} (2018),  030.

\bibitem{x3} M. Blagojevi\'c, \emph{Gravitation and Gauge Symmetries}
    (IOP Publishing, Bristol, 2002);

    T. Ort\'{\i}n, \emph{Gravity and Strings} (Cambridge University Press,
    Cambridge, United Kingdom, 2004).
\bibitem{x4} Yu. N. Obukhov, Poincar\'e gauge gravity: Selected topics,
    Int. J. Geom. Meth. Mod. Phys. {\bf 03}, 95 (2006).

\bibitem{x5} \emph{Gauge Theories
    of Gravitation, A Reader with Commentaries}, edited by M. Blagojevi\'c and F. W. Hehl,  (Imperial College Press,
    London, 2013).
\bibitem{9} M. Blagojevi\'c and B. Cvetkovi\'c, 3D gravity with
    propagating torsion: The AdS sector, Phys. Rev. D {\bf 85}, 104003 (2012).

\bibitem{10} M. Blagojevi\'c and B. Cvetkovi\'c, Gravitational waves
    with torsion in 3D, Phys. Rev. D {\bf 90}, 044006 (2014).

\bibitem{x23} F. W. Hehl, J. D. McCrea, E. W. Mielke, and Y. Neeman,
    Metric-affine gauge theory of gravity: Field equations, Noether
    identities, world spinors, and breaking of dilation invariance, Phys.
    Rep. {\bf 258}, 1 (1995).

\bibitem{y1} D. Puetzfeld and  Yuri N. Obukhov, Probing non-Riemannian spacetime geometry, Phys. Lett A {\bf 372} 6711 (2008).

\bibitem{y2} P. Baekler, A spherically symmetric vacuum solution of the quadratic Poincar\'e gauge
field theory of gravitation with Newtonian and confinement potentials, Phys. Lett. {\bf 99}
B (1981) 329-332;

J. D. McCrea, P. Baekler and M. G\"urses, A Kerr-like solution of the Poincar\'e gauge
field equations, Nuovo Cimento B {\bf 99} 171 (1987);

\bibitem{y3} Yu. N. Obukhov, Exact solutions in Poincar\'e gauge gravity theory, Universe {\bf 5, 127} (2019).

\bibitem{x19}  G. E. Andrews, R. Askey, and R. Roy, \emph{Special
    Functions} (Cambridge University Press, United Kingdom, Cambridge, 1999);

    Z. X. Wang and D. R. Guo, \emph{Special Functions} (World Scientific,
    Singapore, 1989).

\bibitem{17} E. Kilicarslan and B. Tekin, Graviton mass and memory, Eur. Phys. J. C {\bf 79} 114 (2019).

\bibitem{y4} M. Blagojevi\'c  and B. Cvetkovi\'c, Generalized pp waves in Poincar\'e gauge theory,  Phys. Rev. D {\bf 95}, 104018  (2017);

M. Blagojevi\'c, B. Cvetkovi\'c and Y. N. Obukhov, Generalized plane waves in Poincar\'e gauge theory of gravity, Phys. Rev. D {\bf 96}, 064031  (2017).
\end{thebibliography}
\end{document}